\documentclass[prd,nofootinbib,preprint]{revtex4}
\usepackage{latexsym}

\newcommand{\FORTRAN}{\texttt{FORTRAN}}
\newcommand{\HERWIG}{\texttt{HERWIG}}
\newcommand{\PYTHIA}{\texttt{PYTHIA}}
\newcommand{\POLINT}{\texttt{POLINT}}
\newcommand{\POLINTT}{\texttt{POLINT3}}
\newcommand{\ifc}{\texttt{ifc}}
\newcommand{\gnu}{\texttt{g77}}

\begin{document}


\title{Faster Parton Distribution Evaluation in Monte Carlos}

\author{Z.~Sullivan}
\affiliation{Theoretical Physics Department, Fermi National Accelerator
Laboratory,\\ Batavia, IL 60510-0500, USA}

\date{March 4, 2004}

\begin{abstract}
I recommend a few trivial changes to the routines that evaluate CTEQ
parton distribution functions.  These changes allow modern compilers
to optimize the evaluation routines, while having no quantitative
effect on the results.  Computation time is reduced by a factor of 2
in matrix-element calculations, and by 1.3--1.5 in the showering Monte
Carlo event generators with fast detector effects included.  The
results suggest that additional time should be invested in optimizing
these routines.
\end{abstract}


\maketitle

\section{Introduction}
\label{sec:introduction}

A significant amount of time and computing resources are spent on
calculating events at hadron colliders.  Whether a theoretical
calculation of matrix elements, or an experimental simulation of
events with detector effects, one common element is the evaluation of
parton distribution functions (PDFs).  In profiling \FORTRAN\ versions
of code \cite{Harris:2002md, ZSnew} written to simulate
single-top-quark production, it has become apparent that much of the
execution time of real production code is spent acquiring these PDFs.
Upon close examination of the CTEQ PDFs \cite{CTEQ}, a handful trivial
optimizations arise that can cut this time in half.  I list specific
recommendations that are simple to implement, but which can have large
consequences for efficiency.

Profiling a \FORTRAN\ version of the code for fully differential
next-to-leading order single-top-quark production \cite{Harris:2002md,
ZSnew} indicates that at least 75\% of the running time is spent in
acquiring PDFs.  Furthermore, most of that time is spent inside the
subroutine \POLINT.  \POLINT\ is a routine designed to perform a
polynomial fit of degree $n-1$ to a data set of $n$ points based on
\emph{Neville's algorithm} \cite{numrec}.  This subroutine is used by
the CTEQ Collaboration \cite{CTEQ} to interpolate smoothly between
the values of $x$ and $Q^2$ that are read in from a table of best-fit
values.  Clearly any increase in speed in evaluating \POLINT\ will
translate directly into decreased execution time for the program.

One approach to increasing speed would be to replace \POLINT\ outright
with alternate interpolations, or functional fits to the PDFs.  While
these are reasonable choices, I wish to retain the same method to
ensure that any results are numerically identical to results obtained
previously.  Therefore, I will limit my recommendations to trivial
modifications of \POLINT\ itself.  Two useful changes to the coding are:
\begin{enumerate}
\item Remove the line: \texttt{IF(DEN.EQ.0.)PAUSE}.
\item Write different versions of \POLINT\ for 3- and 4-point interpolation,
and call them directly.  E.g., replace \texttt{POLINT(XA,YA,N,X,Y,DY)}
with \texttt{POLINT3(XA,YA,3,X,Y,DY)}, and explicitly set \texttt{N} to 3 in
the code.
\end{enumerate}
The first optimization is the most important as the line is never
reached in the evaluation of the CTEQ PDFs, but it generally prevents
the compiler from fully optimizing the loops.  Beyond being an
unnecessary comparison, the real problem is that allowing a break
point out of the loop can disallow some types of parallel instructions
to be passed to the CPU.  The second optimization mostly helps
compilers to optimize the loops by defining the number of iterations
at compile time, rather than dynamically.

\section{Optimization results}
\label{sec:results}

In order to assess the usefulness of these optimizations, I evaluate
four programs.  I consider a loop over PDFs, and three calculations of
$t$-channel single-top-quark production: an analytic next-to-leading
order calculation of jet distributions \cite{Harris:2002md, ZSnew},
\HERWIG\ 6.1 \cite{Corcella:2000bw}, and \PYTHIA\ 6.2
\cite{Sjostrand:2000wi}.  The first two calculations are compiled
with the GNU compiler \gnu\ versions 2.95 and 3.1, and the Intel
compiler \ifc\ 6.0.  The two showering event generators have only been
considered with the Intel compiler because they were linked with a
fast detector simulation called \texttt{SHW} \cite{Carena:2000yx} that
requires additional libraries that would also have to be recompiled.
While additional information might be gained by trying more compilers,
there are program specific issues that would also have to be
addressed.  All numerical results are from execution on a 1.4 GHz
Pentium 4 machine.  Limited tests performed on Pentium 3 machines are
completely consistent with the results described below.

\subsection{Benchmark for PDFs}

The most naive test of potential speed gains comes from looping over
parton distribution functions.  For this benchmark, values of $x$ are
looped over, the scale is chosen to be $Q^2=(x*1960 \mathrm{\
GeV})^2$, and all parton flavors are evaluated.  These choices are
both more representative of an actual program, and avoid any
possibility of anomalous gains due to fast memory access in the level
2 cache.  The most immediate observation is that there is less than a
5\% difference between \gnu\ 2.95 and \gnu\ 3.1.\footnote{One
exception is that specific Pentium 4 optimizations in \protect\gnu\
3.1 can increase the difference to 10\%.}  Furthermore, the fastest
time I have been able to achieve using \gnu\ is slower than
unoptimized times I obtain by using \ifc\ 6.0.  Since this is true in
all cases tested, I split the results by compiler.

Using \gnu, the running time ranges from 30--39 s.  Removing
\texttt{IF(DEN.EQ.0.)PAUSE} increases the speed by 10\%.  Using
\POLINTT\ instead of \POLINT\ increases speed by an additional 10\%.  The
cumulative effect is a net 20\% gain (30\% if specialized instructions
for the Pentium 4 are used).

The results for using the Intel compiler are more complicated, and
lead to some general conclusions.  First, the Intel compiler performs
more analysis in generating assembly code and branch prediction.  If
poor choices of optimization flags are made, the code can run a factor
of 2 slower.  If default optimization is chosen, the base time of
execution is 30 s.  Using \POLINTT\ instead of \POLINT\ increases speed by
up to 20\%.  Interestingly, just removing \texttt{IF(DEN.EQ.0.)PAUSE}
from \POLINT\ increases the speed by 26\%, and the combination of both
is 30\% faster.  The fact that the optimization is not strictly
multiplicative indicates there are subtle effects involved in how
instructions and data are fed to the processor.  By trying various
compiler flags, it appears that generally the optimization is all or
nothing.  As we will see next, this artificial test code actually
underestimates the potential speed gain for the Intel compiler.

\subsection{Matrix element codes}

Artificial benchmarks can be misleading.  Therefore, I consider the
effects of both changes on working production coding of
single-top-quark production \cite{Harris:2002md, ZSnew}.  The results
using the \gnu\ compiler are identical to the benchmark scenario above
(10\% increases in speed for each change, and if Pentium 4
specialization is used).  With a net 30\% improvement in speed, the
program can run in 80 s when compiled with \gnu.  This is a factor
of 1.4 faster than without the changes.

The Intel compiler provides more interesting results.  The default
running time is 83 s.  Typically there is a 20\% increase in speed
when using \POLINTT\ instead of \POLINT.  Dropping
\texttt{IF(DEN.EQ.0.)PAUSE} alone increases the speed by 33\%; and
both optimizations together yield a net speed increase of 35\%
(execution times of about 54 s).  It is interesting that the most
efficient compiler options for the optimized code are actually 45\%
(1.8 times) faster than when using the same options in the original
code (which runs in 98 s).  It appears that adding both optimizations
reduces the dependence on compiler options of the net execution time.
Hence, a combination of both optimizations seems a desirable option.

Since matrix element calculations spend so much time calling PDFs it is
prudent to make one additional recommendation:
\begin{enumerate}
\item[3.] Eliminate any unnecessary calls to the PDFs.
\end{enumerate}
While this may seem obvious, it can be less simple to implement in
practice.  Both \HERWIG\ and \PYTHIA\ use the \texttt{PDFLIB}
\cite{CERNLIB} interface \texttt{STRUCTM} from \texttt{CERNLIB} to
access PDFs.  The \texttt{PDFLIB} routines return a full (though not
necessarily complete) set of unique PDFs --- $g$, $u$, $d$, $\bar u$,
$\bar d$, $s$, $c$, and $b$.  In some matrix elements, not all PDFs
are always needed.  In particular, the leading order diagram for
$t$-channel single-top-quark production requires only the $b$ or $\bar
b$ PDFs from one of the incoming hadrons at a time.  By eliminating
calls to PDFs that are never used, execution time is cut by an
additional 33\%.  Hence the net code is actually up to a \emph{factor
of 2.6} faster than by default.  If there is a clear way to eliminate
extraneous calls to PDFs when coding matrix elements, it should be
implemented.

\subsection{\HERWIG\ and \PYTHIA}

While many theoretical calculations are still only at the
matrix-element level, experiments and careful phenomenological studies
generally resort to using showering Monte Carlo event generators.
These codes are significantly more complex, and we might expect to see
less gain in efficiency as time is spent in showering, and detector
simulation.  In order to assess the impact on the two most common
event generators, \HERWIG\ \cite{Corcella:2000bw} and \PYTHIA\
\cite{Sjostrand:2000wi}, I use them to calculate $t$-channel
single-top-quark production, including all showering effects, etc.  I
also run the output through the fast detector simulation \texttt{SHW}
\cite{Carena:2000yx} in order to represent the most complicated
calculation a phenomenologist is likely to perform.

The \HERWIG\ event generator appears to spend just as much time
evaluating PDFs as the matrix element example considered above.  Using
\POLINTT\ instead of \POLINT\ improves the speed by 20\%.  Removing
\texttt{IF(DEN.EQ.0.)PAUSE} alone improves the speed by 22\%.  As in
the matrix-element case, adding both optimizations to the Intel \ifc\
6.0 compiler results in a combined gain of 25\% (1.33 times faster).
It is interesting to note that the gain can be as large as 33\% (1.5
times faster) if non-optimal compiler flags are chosen, as is often
the case when linking against precompiled libraries.  If both
recommended optimizations are used, however, the actual time of
execution is less sensitive to compiling options.

The improvement in \PYTHIA\ is not as large as in \HERWIG, but it is
still significant.  Removing \texttt{IF(DEN.EQ.0.)PAUSE} or using
\POLINTT\ independently reduce the time of computation by 15\%
(about 1.2 times faster).  Including both optimizations appears to
provide only a percent or two in further improvement.  \PYTHIA\
provides internal routines for some CTEQ PDFs based on functional
fits.  By implementing these changes to \POLINT, the more accurate
table-based evaluation is actually a few percent faster than the
supposedly optimized, but less accurate fits.  This suggests that
parameterizations may not really be needed any longer.  At very least
some profiling should be done before it is determined that hand-coded
optimizations should be used over more general results.

\section{Conclusions}

Given recent improvements in compilers, it behooves us to reconsider
where bottlenecks in computational speed arise.  It appears that one
source of significant loss of computational speed is in evaluating
parton distribution functions.  For users of the CTEQ PDFs I propose
making three simple changes to the code that can cut total
computational time by a factor of 2 or more.  First, I recommend that
unnecessary calls to the PDFs be removed.  In the case of interface
functions the $\bar{c}$, and $\bar{b}$ PDFs may directly set equal to
the already evaluated $c$, and $b$ PDFs, respectively.  This is
already done in the \texttt{PDFLIB} functions \texttt{STRUCTM} and
\texttt{PFTOPDG}.  Note that the \texttt{PDFLIB} routines also set
$\bar{s}=s$, which is an interface bug for \texttt{PFTOPDG}, since
newer PDFs may not have the same value for both partons.  Care must be
taken to ensure that a given set of PDFs are consistently called.

The second recommendation is to make 2 trivial changes to the
subroutine \POLINT.  Comment out the line
\texttt{IF(DEN.EQ.0.)PAUSE}, and replace the general $n$-point fitting
\POLINT\ with versions specialized to 3- or 4-point fits.  Depending on
the exact compiler and program, gains of 20--30\% (factors of
1.25--1.4) are typical with, with gains up to a factor of 2
possible if the right compiler optimizations are chosen.  These
routines should be changed in the base CTEQ distribution \cite{CTEQ},
the \texttt{PDFLIB} routines in \texttt{CERNLIB} \cite{CERNLIB}, and
in the new Les Houches Accord compilation of PDFs \texttt{LHAPDF}
\cite{Giele:2002hx}.

Despite the large gains in speed obtained by the changes I propose,
many programs will still spend most of their time calling PDF
subroutines.  This suggests two paths that should be followed.  First,
a systematic study of the CTEQ evolution code should be performed to
determine whether there are places that the code could be rewritten to
improve efficiency without changing the numerical output.  This would
allow universal improvements in code execution.  Second, each Monte
Carlo writer should be aware of these timing issues (and potential
bugs if $\bar s$ is not the same as $s$), and attempt to reduce
unnecessary calls to the PDFs in their own code.  Finally, it is
interesting that these optimizations can remove the need for less
accurate parameterizations.

\begin{acknowledgments}
This work was supported by the U.~S.~Department of Energy under
Contract No.\ DE-AC02-76CH03000.
\end{acknowledgments}

\end{document}